\PassOptionsToPackage{unicode}{hyperref}
\PassOptionsToPackage{hyphens}{url}
\documentclass[
]{article}
\usepackage{amsmath,amssymb}
\usepackage{lmodern}
\usepackage{ifxetex,ifluatex}
\ifnum 0\ifxetex 1\fi\ifluatex 1\fi=0 
  \usepackage[T1]{fontenc}
  \usepackage[utf8]{inputenc}
  \usepackage{textcomp} 
\else 
  \usepackage{unicode-math}
  \defaultfontfeatures{Scale=MatchLowercase}
  \defaultfontfeatures[\rmfamily]{Ligatures=TeX,Scale=1}
\fi
\IfFileExists{upquote.sty}{\usepackage{upquote}}{}
\IfFileExists{microtype.sty}{
  \usepackage[]{microtype}
  \UseMicrotypeSet[protrusion]{basicmath} 
}{}
\makeatletter
\@ifundefined{KOMAClassName}{
  \IfFileExists{parskip.sty}{%
    \usepackage{parskip}
  }{
    \setlength{\parindent}{0pt}
    \setlength{\parskip}{6pt plus 2pt minus 1pt}}
}{
  \KOMAoptions{parskip=half}}
\makeatother
\usepackage{xcolor}
\IfFileExists{xurl.sty}{\usepackage{xurl}}{} 
\IfFileExists{bookmark.sty}{\usepackage{bookmark}}{\usepackage{hyperref}}
\hypersetup{
  pdftitle={Trade-off between deep learning for species identification and inference about predator-prey co-occurrence: Reproducible R workflow integrating models in computer vision and ecological statistics},
  pdfauthor={O. Gimenez, M. Kervellec, J.-B. Fanjul, A. Chaine, L. Marescot, Y. Bollet, C. Duchamp},
  hidelinks,
  pdfcreator={LaTeX via pandoc}}
\usepackage[margin=1in]{geometry}
\usepackage{color}
\usepackage{fancyvrb}

\DefineVerbatimEnvironment{Highlighting}{Verbatim}{commandchars=\\\{\}}
\usepackage{framed}
\definecolor{shadecolor}{RGB}{248,248,248}
\newenvironment{Shaded}{\begin{snugshade}}{\end{snugshade}}

\newcommand{\AttributeTok}[1]{\textcolor[rgb]{0.77,0.63,0.00}{#1}}

\newcommand{\CommentTok}[1]{\textcolor[rgb]{0.56,0.35,0.01}{\textit{#1}}}

\newcommand{\ConstantTok}[1]{\textcolor[rgb]{0.00,0.00,0.00}{#1}}
\newcommand{\ControlFlowTok}[1]{\textcolor[rgb]{0.13,0.29,0.53}{\textbf{#1}}}

\newcommand{\DecValTok}[1]{\textcolor[rgb]{0.00,0.00,0.81}{#1}}

\newcommand{\FloatTok}[1]{\textcolor[rgb]{0.00,0.00,0.81}{#1}}
\newcommand{\FunctionTok}[1]{\textcolor[rgb]{0.00,0.00,0.00}{#1}}

\newcommand{\NormalTok}[1]{#1}

\newcommand{\OtherTok}[1]{\textcolor[rgb]{0.56,0.35,0.01}{#1}}

\newcommand{\SpecialCharTok}[1]{\textcolor[rgb]{0.00,0.00,0.00}{#1}}

\newcommand{\StringTok}[1]{\textcolor[rgb]{0.31,0.60,0.02}{#1}}

\usepackage{graphicx}
\makeatletter
\def\maxwidth{\ifdim\Gin@nat@width>\linewidth\linewidth\else\Gin@nat@width\fi}
\def\maxheight{\ifdim\Gin@nat@height>\textheight\textheight\else\Gin@nat@height\fi}
\makeatother
\setkeys{Gin}{width=\maxwidth,height=\maxheight,keepaspectratio}
\makeatletter
\def\fps@figure{htbp}
\makeatother
\usepackage{booktabs}
\usepackage{longtable}
\usepackage{array}
\usepackage{multirow}
\usepackage{wrapfig}
\usepackage{float}
\usepackage{colortbl}
\usepackage{pdflscape}
\usepackage{tabu}
\usepackage{threeparttable}
\usepackage{threeparttablex}
\usepackage[normalem]{ulem}
\usepackage{makecell}
\usepackage{xcolor}
\ifluatex
  \usepackage{selnolig}  
\fi
\newlength{\cslhangindent}
\newlength{\csllabelwidth}
\newenvironment{CSLReferences}[2] 
 {
  \setlength{\parindent}{0pt}
  \ifodd #1 \everypar{\setlength{\hangindent}{\cslhangindent}}\ignorespaces\fi
  \ifnum #2 > 0
  \setlength{\parskip}{#2\baselineskip}
  \fi
 }%
 {}
\usepackage{calc}

\title{Trade-off between deep learning for species identification and
inference about predator-prey co-occurrence: Reproducible \texttt{R}
workflow integrating models in computer vision and ecological
statistics}
\author{O. Gimenez, M. Kervellec, J.-B. Fanjul, A. Chaine, L. Marescot,
Y. Bollet, C. Duchamp}
\date{2021-08-26}

\begin{document}
\maketitle

\hypertarget{abstract}{%
\section{Abstract}\label{abstract}}

Deep learning is used in computer vision problems with important
applications in several scientific fields. In ecology for example, there
is a growing interest in deep learning for automatizing repetitive
analyses on large amounts of images, such as animal species
identification.

However, there are challenging issues toward the wide adoption of deep
learning by the community of ecologists. First, there is a programming
barrier as most algorithms are written in \texttt{Python} while most
ecologists are versed in \texttt{R}. Second, recent applications of deep
learning in ecology have focused on computational aspects and simple
tasks without addressing the underlying ecological questions or carrying
out the statistical data analysis to answer these questions.

Here, we showcase a reproducible \texttt{R} workflow integrating both
deep learning and statistical models using predator-prey relationships
as a case study. We illustrate deep learning for the identification of
animal species on images collected with camera traps, and quantify
spatial co-occurrence using multispecies occupancy models.

Despite average model classification performances, ecological inference
was similar whether we analysed the ground truth dataset or the
classified dataset. This result calls for further work on the trade-offs
between time and resources allocated to train models with deep learning
and our ability to properly address key ecological questions with
biodiversity monitoring. We hope that our reproducible workflow will be
useful to ecologists and applied statisticians.

All material (source of the Rmarkdown notebook and auxiliary files) is
available from
\url{https://github.com/oliviergimenez/computo-deeplearning-occupany-lynx}.

\hypertarget{introduction}{%
\section{Introduction}\label{introduction}}

Computer vision is a field of artificial intelligence in which a machine
is taught how to extract and interpret the content of an image
(\protect\hyperlink{ref-NIPS2012_4824}{Krizhevsky, Sutskever, and Hinton
2012}). Computer vision relies on deep learning that allows
computational models to learn from training data -- a set of manually
labelled images -- and make predictions on new data -- a set of
unlabelled images
(\protect\hyperlink{ref-baraniuk_science_2020}{Baraniuk, Donoho, and
Gavish 2020}; \protect\hyperlink{ref-lecun_deep_2015}{LeCun, Bengio, and
Hinton 2015}). With the growing availability of massive data, computer
vision with deep learning is being increasingly used to perform tasks
such as object detection, face recognition, action and activity
recognition or human pose estimation in fields as diverse as medicine,
robotics, transportation, genomics, sports and agriculture
(\protect\hyperlink{ref-andina_deep_2018}{Voulodimos et al. 2018}).

In ecology in particular, there is a growing interest in deep learning
for automatizing repetitive analyses on large amounts of images, such as
identifying plant and animal species, distinguishing individuals of the
same or different species, counting individuals or detecting relevant
features (\protect\hyperlink{ref-christin_applications_2019}{Christin,
Hervet, and Lecomte 2019}; \protect\hyperlink{ref-lamba_deep_2019}{Lamba
et al. 2019}; \protect\hyperlink{ref-weinstein_computer_2018}{Weinstein
2018}). By saving hours of manual data analyses and tapping into massive
amounts of data that keep accumulating with technological advances, deep
learning has the potential to become an essential tool for ecologists
and applied statisticians.

Despite the promising future of computer vision and deep learning, there
are challenging issues toward their wide adoption by the community of
ecologists (e.g. \protect\hyperlink{ref-wearn_responsible_2019}{Wearn,
Freeman, and Jacoby 2019}). First, there is a programming barrier as
most, if not all, algorithms are written in the \texttt{Python} language
while most ecologists are versed in \texttt{R}
(\protect\hyperlink{ref-lai_evaluating_2019}{Lai et al. 2019}). If
ecologists are to use computer vision in routine, there is a need for
bridges between these two languages (through, e.g., the
\texttt{reticulate} package
\protect\hyperlink{ref-reticulate_ref}{Allaire et al.}
(\protect\hyperlink{ref-reticulate_ref}{2017}) or the \texttt{shiny}
package \protect\hyperlink{ref-tabak_improving_2020}{Tabak et al.}
(\protect\hyperlink{ref-tabak_improving_2020}{2020})). Second, recent
applications of computer vision via deep learning in ecology have
focused on computational aspects and simple tasks without addressing the
underlying ecological questions
(\protect\hyperlink{ref-sutherland_identification_2013}{Sutherland et
al. 2013}), or carrying out statistical data analysis to answer these
questions (\protect\hyperlink{ref-gimenez_statistical_2014}{Gimenez et
al. 2014}). Although perfectly understandable given the challenges at
hand, we argue that a better integration of the \emph{why} (ecological
questions), the \emph{what} (automatically labelled images) and the
\emph{how} (statistics) would be beneficial to computer vision for
ecology (see also
\protect\hyperlink{ref-weinstein_computer_2018}{Weinstein 2018}).

Here, we showcase a full why-what-how workflow in \texttt{R} using a
case study on the structure of an ecological community (a set of
co-occurring species) composed of the Eurasian lynx (\emph{Lynx lynx})
and its two main preys. First, we introduce the case study and motivate
the need for deep learning. Second we illustrate deep learning for the
identification of animal species in large amounts of images, including
model training and validation with a dataset of labelled images, and
prediction with a new dataset of unlabelled images. Last, we proceed
with the quantification of spatial co-occurrence using statistical
models.

\hypertarget{collecting-images-with-camera-traps}{%
\section{Collecting images with camera
traps}\label{collecting-images-with-camera-traps}}

Lynx (\emph{Lynx lynx}) went extinct in France at the end of the 19th
century due to habitat degradation, human persecution and decrease in
prey availability
(\protect\hyperlink{ref-vandel_distribution_2005}{Vandel and Stahl
2005}). The species was reintroduced in Switzerland in the 1970s
(\protect\hyperlink{ref-breitenmoser_large_1998}{Breitenmoser 1998}),
then re-colonised France through the Jura mountains in the 1980s
(\protect\hyperlink{ref-vandel_distribution_2005}{Vandel and Stahl
2005}). The species is listed as endangered under the 2017 IUCN Red list
and is of conservation concern in France due to habitat fragmentation,
poaching and collisions with vehicles. The Jura holds the bulk of the
French lynx population.

To better understand its distribution, we need to quantify its
interactions with its main preys, roe deer (\emph{Capreolus capreolus})
and chamois (\emph{Rupicapra rupicapra})
(\protect\hyperlink{ref-molinari-jobin_variation_2007}{Molinari-Jobin et
al. 2007}), two ungulate species that are also hunted. To assess the
relative contribution of predation and hunting, a predator-prey program
was set up jointly by the French Office for Biodiversity, the
Federations of Hunters from the Jura, Ain and Haute-Savoie counties and
the French National Centre for Scientific Research.

Animal detections were made using a set of camera traps in the Jura
mountains that were deployed in the Jura and Ain counties (see Figure
1). We divided the two study areas into grids of 2.7 \(\times\) 2.7 km
cells or sites hereafter
(\protect\hyperlink{ref-zimmermann_optimizing_2013}{Zimmermann et al.
2013}) in which we set two camera traps per site (Xenon white flash with
passive infrared trigger mechanisms, model Capture, Ambush and Attack;
Cuddeback), with 18 sites in the Jura study area, and 11 in the Ain
study area that were active over the study period (from February 2016 to
October 2017 for the Jura county, and from February 2017 to May 2019 for
the Ain county). The location of camera traps was chosen to maximise
lynx detection. Camera traps were checked weekly to change memory cards,
batteries and to remove fresh snow after heavy snowfall.

\begin{figure}
\centering
\includegraphics{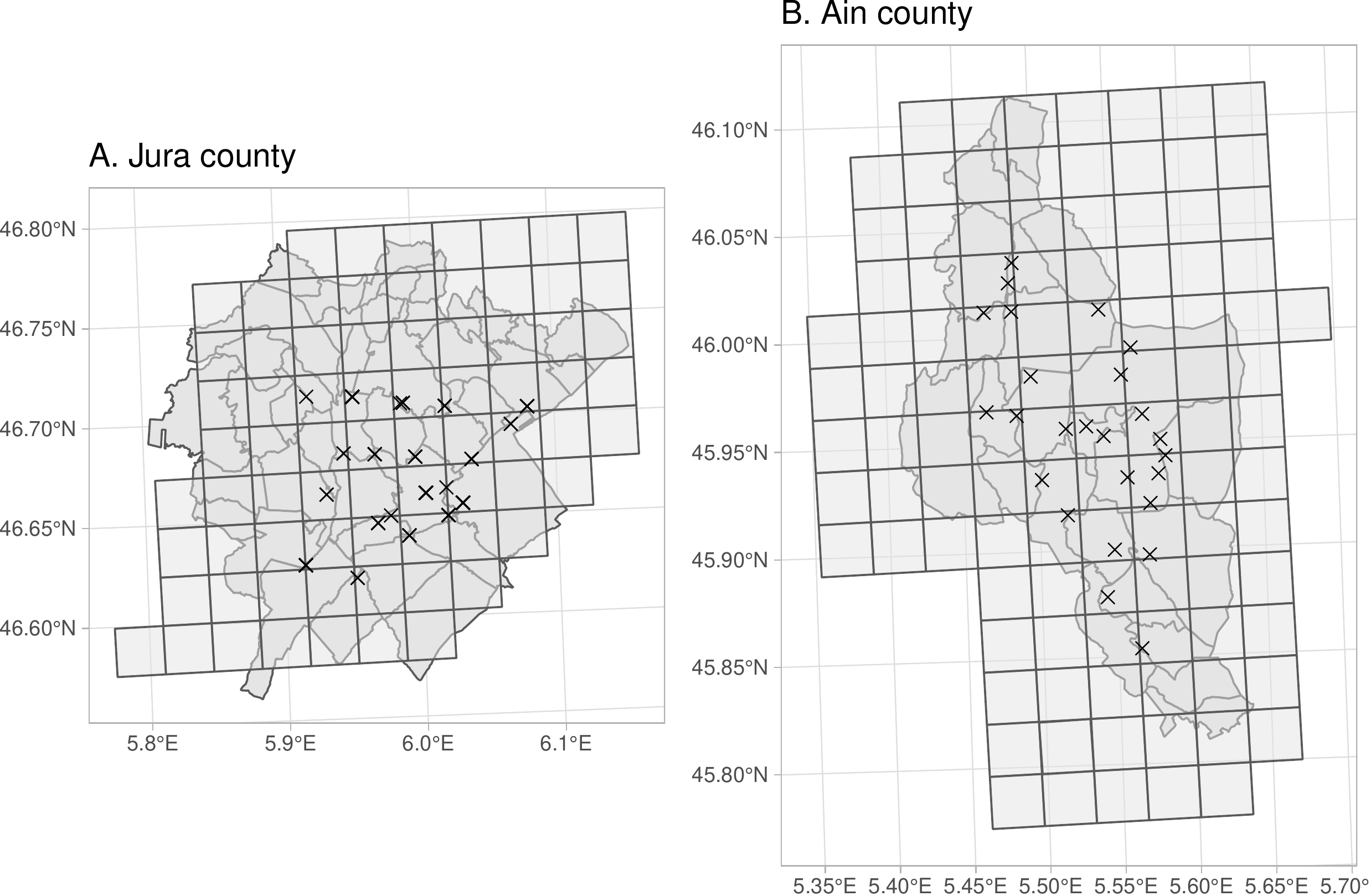}
\caption{Study area, grid and camera trap locations.}
\end{figure}

In total, 45563 and 18044 pictures were considered in the Jura and Ain
sites respectively after manually droping empty pictures and pictures
with unidentified species. Note that classifying empty images could be
automatised with deep learning
(\protect\hyperlink{ref-norouzzadeh_deep_2021}{Norouzzadeh et al. 2021};
\protect\hyperlink{ref-tabak_improving_2020}{Tabak et al. 2020}). We
identified the species present on all images by hand (see Table 1) using
\texttt{digiKam} a free open-source digital photo management application
(\url{https://www.digikam.org/}). This operation took several weeks of
labor full time, which is often identified as a limitation of camera
trap studies. To expedite this tedious task, computer vision with deep
learning has been identified as a promising approach
(\protect\hyperlink{ref-norouzzadeh_deep_2021}{Norouzzadeh et al. 2021};
\protect\hyperlink{ref-tabak_machine_2019}{Tabak et al. 2019};
\protect\hyperlink{ref-willi_identifying_2019}{Willi et al. 2019}).

\begin{table}

\caption{\label{tab:categories}Species identified in the Jura and Ain study sites with samples size (n). Only first 10 species with most images are shown.}
\centering
\begin{tabular}[t]{l|r|l|r}
\hline
Species in Jura study site & n & Species in Ain study site & n\\
\hline
human & 31644 & human & 4946\\
\hline
vehicule & 5637 & vehicule & 4454\\
\hline
dog & 2779 & dog & 2310\\
\hline
fox & 2088 & fox & 1587\\
\hline
chamois & 919 & rider & 1025\\
\hline
wild board & 522 & roe deer & 860\\
\hline
badger & 401 & chamois & 780\\
\hline
roe deer & 368 & hunter & 593\\
\hline
cat & 343 & wild board & 514\\
\hline
lynx & 302 & badger & 461\\
\hline
\end{tabular}
\end{table}

\hypertarget{deep-learning-for-species-identification}{%
\section{Deep learning for species
identification}\label{deep-learning-for-species-identification}}

Using the images we obtained with camera traps (Table 1), we trained a
model for identifying species using the Jura study site as a calibration
dataset. We then assessed this model's ability to automatically identify
species on a new dataset, also known as transferability, using the Ain
study site as an evaluation dataset.

\hypertarget{training---jura-study-site}{%
\subsection{Training - Jura study
site}\label{training---jura-study-site}}

We selected at random 80\% of the annotated images for each species in
the Jura study site for training, and 20\% for testing. We applied
various transformations (flipping, brightness and contrast
modifications; \protect\hyperlink{ref-shorten_survey_2019}{Shorten and
Khoshgoftaar} (\protect\hyperlink{ref-shorten_survey_2019}{2019})) to
improve training (see Appendix). To reduce model training time and
overcome the small number of images, we used transfer learning
(\protect\hyperlink{ref-Yosinski2014}{Yosinski et al. 2014};
\protect\hyperlink{ref-shao_transfer_2015}{Shao, Zhu, and Li 2015}) and
considered a pre-trained model as a starting point. Specifically, we
trained a deep convolutional neural network (ResNet-50) architecture
(\protect\hyperlink{ref-he_deep_2016}{He et al. 2016}) using the
\texttt{fastai} library (\url{https://docs.fast.ai/}) that implements
the \texttt{PyTorch} library
(\protect\hyperlink{ref-NEURIPS2019_9015}{Paszke et al. 2019}).
Interestingly, the \texttt{fastai} library comes with an \texttt{R}
interface (\url{https://eagerai.github.io/fastai/}) that uses the
\texttt{reticulate} package to communicate with \texttt{Python},
therefore allowing \texttt{R} users to access up-to-date deep learning
tools. We trained models on the Montpellier Bioinformatics Biodiversity
platform using a GPU machine (Titan Xp nvidia) with 16Go of RAM. We used
20 epochs which took approximately 10 hours. The computational burden
prevented us from providing a full reproducible analysis, but we do so
with a subsample of the dataset in the Appendix. All trained models are
available from \url{https://doi.org/10.5281/zenodo.5164796}.

We calculated three metrics to evaluate our model performance at
correctly identifying species (e.g.
\protect\hyperlink{ref-Duggan2021}{Duggan et al. 2021}). Specifically,
we relied on \emph{accuracy} the ratio of correct predictions to the
total number of predictions, \emph{recall} a measure of false negatives
(FN; e.g.~an image with a lynx for which our model predicts another
species) with recall = TP / (TP + FN) where TP is for true positives,
and \emph{precision} a measure of false positives (FP; e.g.~an image
with any species but a lynx for which our model predicts a lynx) with
precision = TP / (TP + FP). In camera trap studies, a strategy
(\protect\hyperlink{ref-Duggan2021}{Duggan et al. 2021}) consists in
optimizing precision if the focus is on rare species (lynx), while
recall should be optimized if the focus is on commom species (chamois
and roe deer).

We achieved 85\% accuracy during training. Our model had good
performances for the three classes we were interested in, with 87\%
precision for lynx and 81\% recall for both roe deer and chamois (Table
2).

\begin{table}

\caption{\label{tab:perf}Model training performance. Images from the Jura study site were used for training.}
\centering
\begin{tabular}[t]{>{}l|l|l}
\hline
species & precision & recall\\
\hline
\textbf{badger} & 0.78 & 0.88\\
\hline
\textbf{red deer} & 0.67 & 0.21\\
\hline
\textbf{chamois} & 0.86 & 0.81\\
\hline
\textbf{cat} & 0.89 & 0.78\\
\hline
\textbf{roe deer} & 0.67 & 0.81\\
\hline
\textbf{dog} & 0.78 & 0.84\\
\hline
\textbf{human} & 0.99 & 0.79\\
\hline
\textbf{hare} & 0.32 & 0.52\\
\hline
\textbf{lynx} & 0.87 & 0.95\\
\hline
\textbf{fox} & 0.85 & 0.90\\
\hline
\textbf{wild boar} & 0.93 & 0.88\\
\hline
\textbf{vehicule} & 0.95 & 0.98\\
\hline
\end{tabular}
\end{table}

\hypertarget{transferability---ain-study-site}{%
\subsection{Transferability - Ain study
site}\label{transferability---ain-study-site}}

We evaluated transferability for our trained model by predicting species
on images from the Ain study site which were not used for training.
Precision was 77\% for lynx, and while we achieved 86\% recall for roe
deer, our model performed poorly for chamois with 8\% recall (Table 3).

\begin{table}

\caption{\label{tab:transf}Model transferability performance. Images from the Ain study site were used for assessing transferability.}
\centering
\begin{tabular}[t]{>{}l|r|r}
\hline
  & precision & recall\\
\hline
\textbf{badger} & 0.71 & 0.89\\
\hline
\textbf{rider} & 0.79 & 0.92\\
\hline
\textbf{red deer} & 0.00 & 0.00\\
\hline
\textbf{chamois} & 0.82 & 0.08\\
\hline
\textbf{hunter} & 0.17 & 0.11\\
\hline
\textbf{cat} & 0.46 & 0.59\\
\hline
\textbf{roe deer} & 0.67 & 0.86\\
\hline
\textbf{dog} & 0.77 & 0.35\\
\hline
\textbf{human} & 0.51 & 0.93\\
\hline
\textbf{hare} & 0.37 & 0.35\\
\hline
\textbf{lynx} & 0.77 & 0.89\\
\hline
\textbf{marten} & 0.05 & 0.04\\
\hline
\textbf{fox} & 0.90 & 0.53\\
\hline
\textbf{wild board} & 0.75 & 0.94\\
\hline
\textbf{cow} & 0.01 & 0.25\\
\hline
\textbf{vehicule} & 0.94 & 0.51\\
\hline
\end{tabular}
\end{table}

To better understand this pattern, we display the results under the form
of a confusion matrix that compares model classifications to manual
classifications (Figure 2). There were a lot of false negatives for
chamois, meaning that when a chamois was present in an image, it was
often classified as another species by our model.

\begin{figure}
\centering
\includegraphics{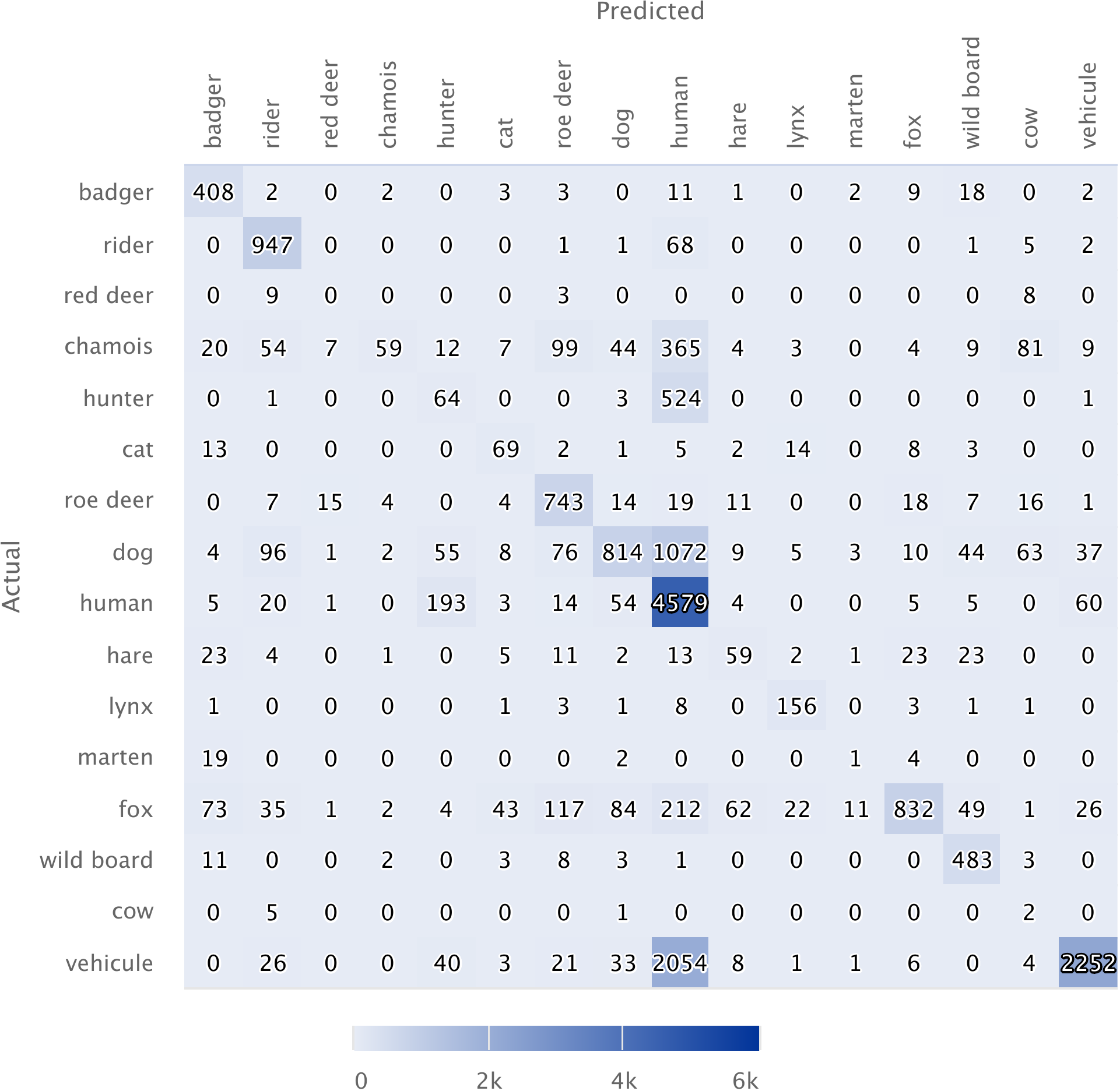}
\caption{Confusion matrix comparing automatic to manual species
classifications. Species that were predicted by our model are in
columns, and species that are actually in the images are in rows.}
\end{figure}

Overall, our model trained on images from the Jura study site did poorly
at correctly predicting species on images from the Ain study site. This
result does not come as a surprise, as generalizing classification
algorithms to new environments is known to be difficult
(\protect\hyperlink{ref-beery_recognition_2018}{Beery, Horn, and Perona
2018}). While a computer scientist might be disappointed in these
results, an ecologist would probably wonder whether ecological inference
about the interactions between lynx and its prey is biased by these
average performances, a question we address in the next section.

\hypertarget{spatial-co-occurrence}{%
\section{Spatial co-occurrence}\label{spatial-co-occurrence}}

Here, we analysed the data we acquired from the previous section. For
the sake of comparison, we considered two datasets, one made of the
images manually labelled for both the Jura and Ain study sites pooled
together (\emph{ground truth dataset}), and the other in which we pooled
the images that were manually labelled for the Jura study site and the
images that were automatically labelled for the Ain study site using our
trained model (\emph{classified dataset}).

We formatted the data by generating monthly detection histories, that is
a sequence of detections (\(Y_{sit} = 1\)) and non-detections
(\(Y_{sit} = 0\)), for species \(s\) at site \(i\) and sampling occasion
\(t\) (see Figure 3).

\begin{figure}

{\centering \includegraphics{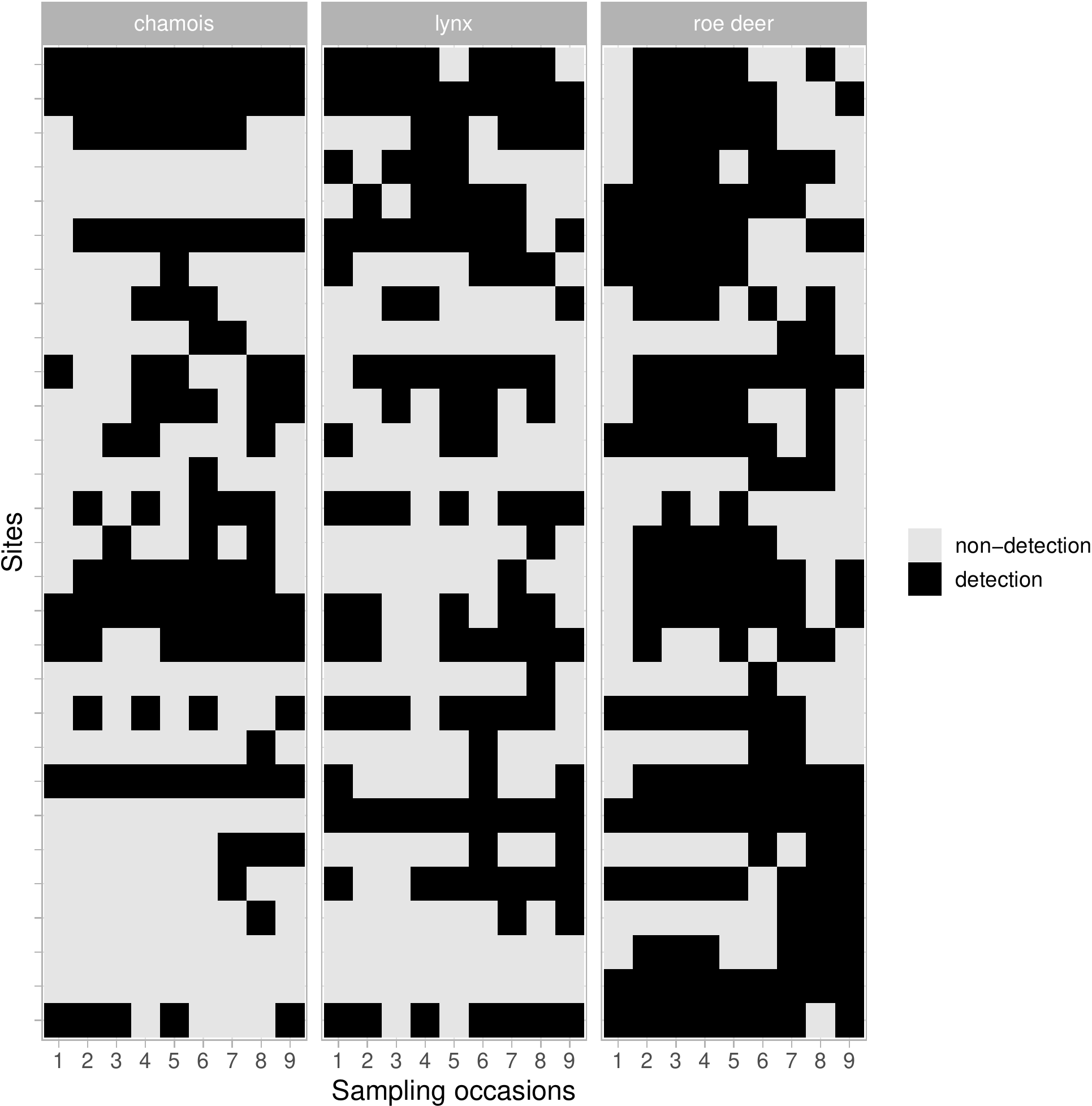} 

}

\caption{Detections (black) and non-detections (light grey) for each of the 3 species lynx, chamois and roe deer between March and November for all years pooled together. Sites are on the Y axis, while sampling occasions are on the X axis. Only data from the ground truth dataset are displayed.}\label{fig:figure3}
\end{figure}

To quantify spatial co-occurrence betwen lynx and its preys, we used a
multispecies occupancy modeling approach
(\protect\hyperlink{ref-Rota2016}{Rota et al. 2016};
\protect\hyperlink{ref-clipp2021}{Clipp et al. 2021}) using the
\texttt{R} package \texttt{unmarked}
(\protect\hyperlink{ref-unmarkedFiske}{Fiske and Chandler 2011}). The
multispecies occupancy model assumes that observations \(y_{sit}\),
conditional on \(Z_{si}\) the latent occupancy state of species \(s\) at
site \(i\) are drawn from Bernoulli random variables
\(Y_{sit} | Z_{si} \sim \mbox{Bernoulli}(Z_{si}p_{sit})\) where
\(p_{sit}\) is the detection probability of species \(s\) at site \(i\)
and sampling occasion \(t\). Detection probabilities can be modeled as a
function of site and/or sampling covariates, or the presence/absence of
other species, but for the sake of illustration, we will make them only
species-specific here.

The latent occupancy states are assumed to be distributed as
multivariate Bernoulli random variables
(\protect\hyperlink{ref-dai_multivariate_2013}{Dai, Ding, and Wahba
2013}). Let us consider 2 species, species 1 and 2, then
\(Z_i = (Z_{i1}, Z_{i2}) \sim \mbox{multivariate Bernoulli}(\psi_{11}, \psi_{10}, \psi_{01}, \psi_{00})\)
where \(\psi_{11}\) is the probability that a site is occupied by both
species 1 and 2, \(\psi_{10}\) the probability that a site is occupied
by species 1 but not 2, \(\psi_{01}\) the probability that a site is
occupied by species 2 but not 1, and \(\psi_{00}\) the probability a
site is occupied by none of them. Note that we considered
species-specific only occupancy probabilities but these could be modeled
as site-specific covariates. Marginal occupancy probabilities are
obtained as \(\Pr(Z_{i1}=1) = \psi_{11} + \psi_{10}\) and
\(\Pr(Z_{i2}=1) = \psi_{11} + \psi_{01}\). With this model, we may also
infer potential interactions by calculating conditional probabilities
such as for example the probability of a site being occupied by species
2 conditional of species 1 with
\(\Pr(Z_{i2} = 1| Z_{i1} = 1) = \displaystyle{\frac{\psi_{11}}{\psi_{11}+\psi_{10}}}\).

Detection probabilities were indistinguishable whether we used the
ground truth or the classified dataset, with
\(p_{\mbox{lynx}} = 0.51 (0.45, 0.58)\),
\(p_{\mbox{roe deer}} = 0.63 (0.57, 0.68)\) and
\(p_{\mbox{chamois}} = 0.61 (0.55, 0.67)\).

We also found that occupancy probability estimates were similar whether
we used the ground truth or the classified dataset (Figure 4). Roe deer
was the prevalent species, but lynx and chamois were also occurring with
high probability (Figure 4).

\begin{figure}
\includegraphics[width=1\linewidth]{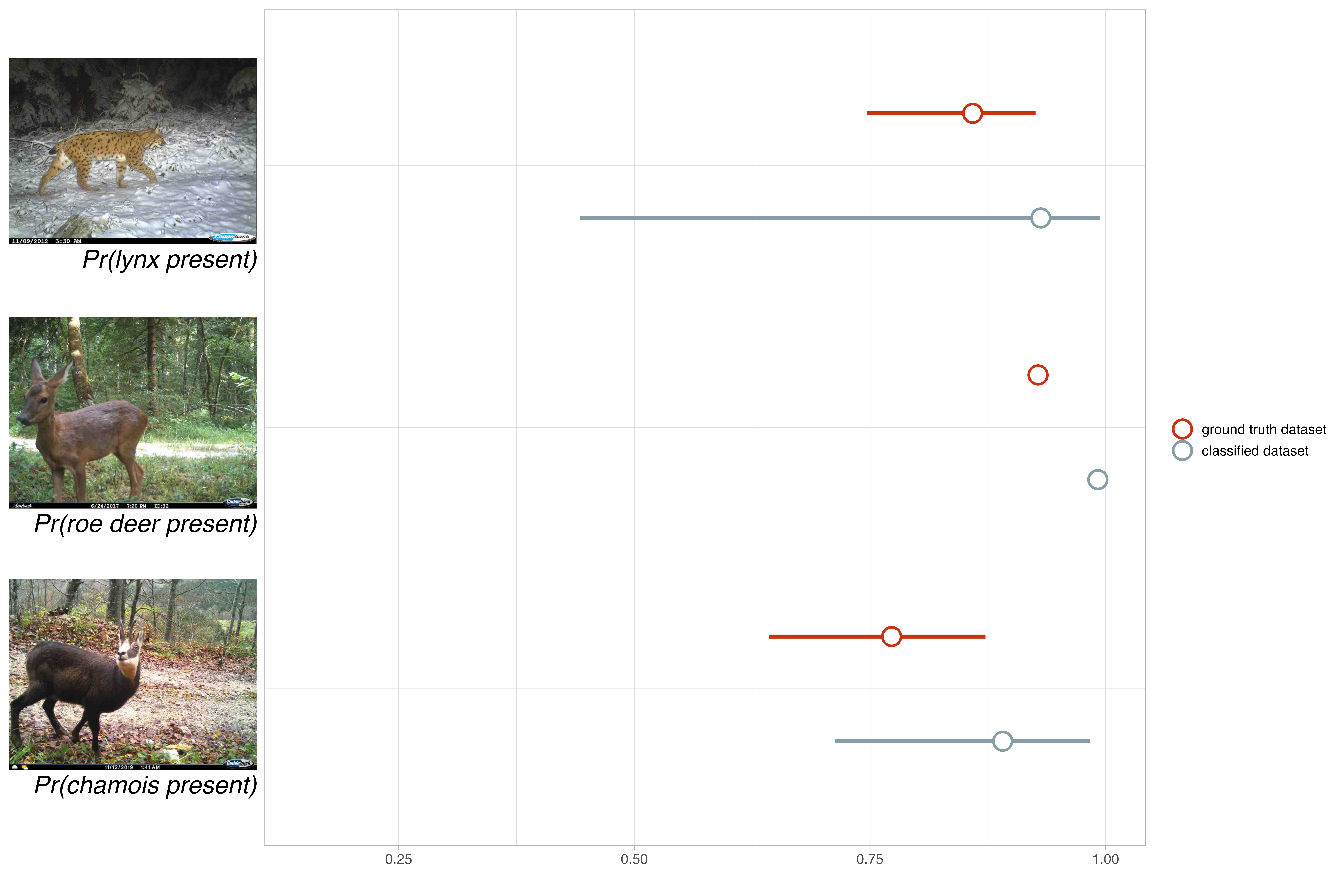} \caption{Marginal occupancy probabilities for all three species, lynx, roe deer and chamois). Parameter estimates are from a multispecies occupancy model using either the ground truth dataset (in red) or the classified dataset (in blue-grey).}\label{fig:figure4}
\end{figure}

Because marginal occupancy probabilities were high, probabilities of
co-occurrence were also estimated high (Figure 5). Our results should be
interpreted bearing in mind that co-occurrence is a necessary but not
sufficient condition for actual interaction. When both preys were
present, lynx was more present than when they were both absent (Figure
5). Lynx was more sensitive to the presence of roe deer than that of
chamois (Figure 5).

\begin{figure}
\includegraphics[width=1\linewidth]{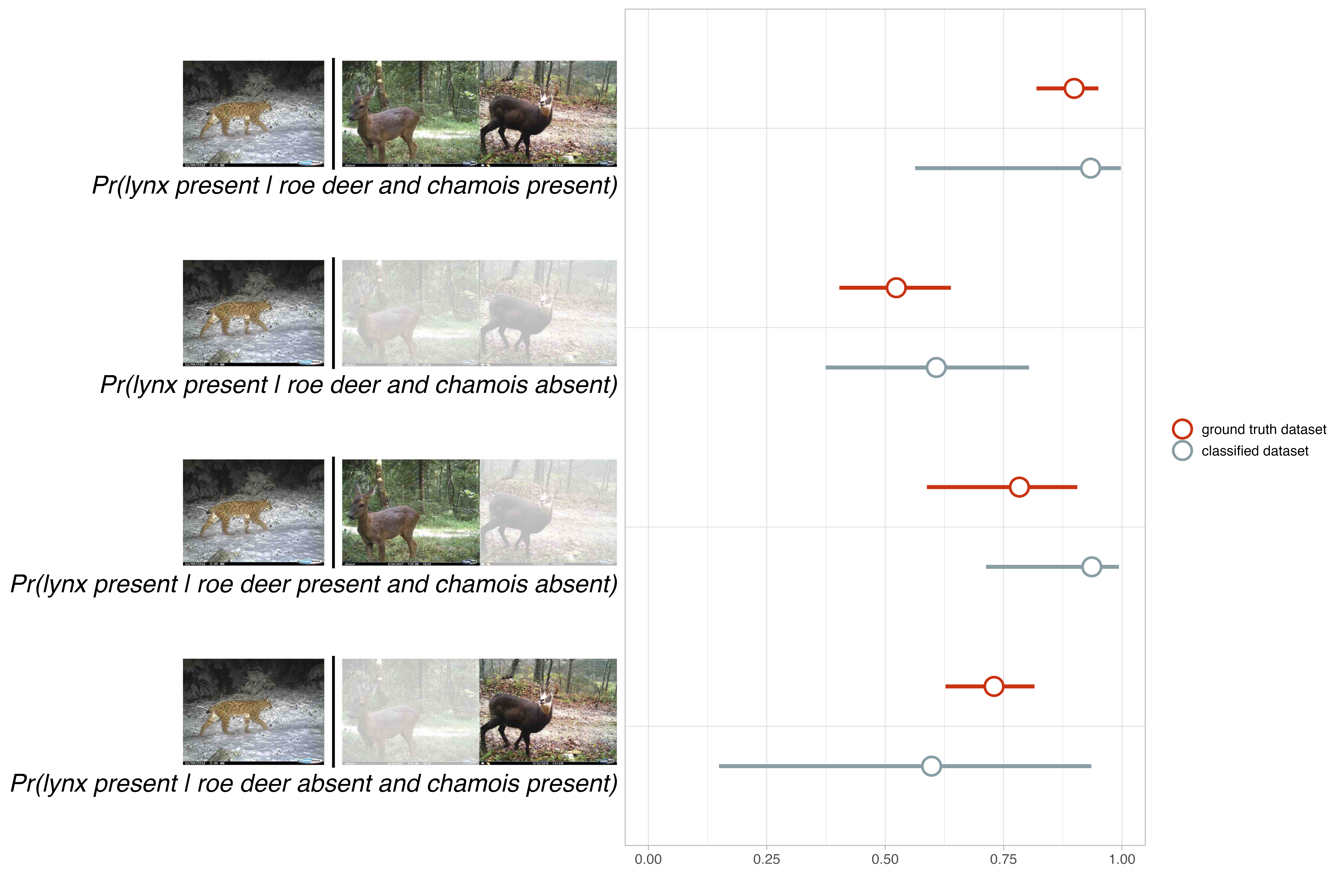} \caption{Lynx occupancy probability conditional on the presence or absence of its preys (roe deer and chamois). Parameter estimates are from a multispecies occupancy model using either the ground truth dataset (in red) or the classified dataset (in blue-grey).}\label{fig:figure5}
\end{figure}

\hypertarget{discussion}{%
\section{Discussion}\label{discussion}}

In this paper, we aimed at illustrating a reproducible workflow for
studying the structure of an animal community and species spatial
co-occurrence (\emph{why}) using images acquired from camera traps and
automatically labelled with deep learning (\emph{what}) which we
analysed with statistical occupancy models accounting for imperfect
species detection (\emph{why}). Overall, we found that, even though
model transferability could be improved, inference about the potential
interactions between lynx and its preys was similar whether we analysed
the ground truth data or classified data.

This result calls for further work on the trade-offs between time and
resources allocated to train models with deep learning and our ability
to correctly answer key ecological questions with camera-trap surveys.
In other words, while a computer scientist might be keen on spending
time training models to achieve top performances, an ecologist would
rather rely on a model showing average performances and use this time to
proceed with statistical analyses if, of course, errors in
computer-annotated images do not make ecological inference flawed. The
right balance may be found with collaborative projects in which
scientists from artificial intelligence, statistics and ecology agree on
a common objective, and identify research questions that can pick the
interest of all parties.

Our demonstration remains however empirical, and we encourage others to
try and replicate our results. We also see two avenues of research that
could benefit the integration of deep learning and ecological
statistics. First, a simulation study could be conducted to evaluate
bias and precision in ecological parameter estimators with regard to
errors in image annotation by computers. The outcome of this exercise
could be, for example, guidelines informing on the confidence an
investigator may place in ecological inference as a function of the
amount of false negatives and false positives. Second, annotation errors
could be accomodated directly in statistical models. For example,
single-species occupancy models account for false negatives when a
species is not detected by the camera at a site where it is present, as
well as false positives when a species is detected at a site where it is
not present due to species misidentification by the observer
(\protect\hyperlink{ref-miller2011}{Miller et al. 2011}). Pending a
careful distinction between ecological vs.~computer-generated false
negatives and false positives, error rates could be added to
multispecies occupancy models
(\protect\hyperlink{ref-chambert2018}{Chambert et al. 2018}) and
informed by recall and precision metrics obtained during model training
(\protect\hyperlink{ref-tabak_improving_2020}{Tabak et al. 2020}).

With regard to the case study, our results should be seen only as
preliminary. First, we aim at quantifying the relative contribution of
biotic (lynx predation on chamois and roe deer) and abiotic (habitat
quality) processes to the composition and dynamic of this ecological
community. Second, to benefit future camera trap studies of lynx in the
Jura mountains, we plan to train a model again using manually annotated
images from both the Jura and the Ain study sites. These perspectives
are the object of ongoing work.

With the rapid advances in technologies for biodiversity monitoring
(\protect\hyperlink{ref-lahoz-monfort_comprehensive_2021}{Lahoz-Monfort
and Magrath 2021}), the possibility of analysing large amounts of images
makes deep learning appealing to ecologists. We hope that our proposal
of a reproducible \texttt{R} workflow for deep learning and statistical
ecology will encourage further studies in the integration of these
disciplines, and contribute to the adoption of computer vision by
ecologists.

\hypertarget{appendix-reproducible-example-of-species-identification-on-camera-trap-images-with-cpu}{%
\section{Appendix: Reproducible example of species identification on
camera trap images with
CPU}\label{appendix-reproducible-example-of-species-identification-on-camera-trap-images-with-cpu}}

In this section, we go through a reproducible example of the entire deep
learning workflow, including data preparation, model training, and
automatic labeling of new images. We used a subsample of 467 images from
the original dataset in the Jura county to allow the training of our
model with CPU on a personal computer. We also used 14 images from the
original dataset in the Ain county to illustrate prediction.

\hypertarget{training-and-validation-datasets}{%
\subsection{Training and validation
datasets}\label{training-and-validation-datasets}}

We first split the dataset of Jura images in two datasets, a dataset for
training, and the other one for validation. We use the \texttt{exifr}
package to extract metadata from images, get a list of images names and
extract the species from these.

\begin{Shaded}
\begin{Highlighting}[]
\FunctionTok{library}\NormalTok{(exifr)}
\NormalTok{pix\_folder }\OtherTok{\textless{}{-}} \StringTok{\textquotesingle{}pix/pixJura/\textquotesingle{}}
\NormalTok{file\_list }\OtherTok{\textless{}{-}} \FunctionTok{list.files}\NormalTok{(}\AttributeTok{path =}\NormalTok{ pix\_folder,}
                        \AttributeTok{recursive =} \ConstantTok{TRUE}\NormalTok{,}
                        \AttributeTok{pattern =} \StringTok{"*.JPG"}\NormalTok{,}
                        \AttributeTok{full.names =} \ConstantTok{TRUE}\NormalTok{)}
\NormalTok{labels }\OtherTok{\textless{}{-}}
  \FunctionTok{read\_exif}\NormalTok{(file\_list) }\SpecialCharTok{\%\textgreater{}\%}
  \FunctionTok{as\_tibble}\NormalTok{() }\SpecialCharTok{\%\textgreater{}\%}
  \FunctionTok{unnest}\NormalTok{(Keywords, }\AttributeTok{keep\_empty =} \ConstantTok{TRUE}\NormalTok{) }\SpecialCharTok{\%\textgreater{}\%} \CommentTok{\# keep\_empty = TRUE keeps pix with no labels (empty pix)}
  \FunctionTok{group\_by}\NormalTok{(SourceFile) }\SpecialCharTok{\%\textgreater{}\%}
  \FunctionTok{slice\_head}\NormalTok{() }\SpecialCharTok{\%\textgreater{}\%} \CommentTok{\# when several labels in a pix, keep first only}
  \FunctionTok{ungroup}\NormalTok{() }\SpecialCharTok{\%\textgreater{}\%}
  \FunctionTok{mutate}\NormalTok{(}\AttributeTok{Keywords =} \FunctionTok{as\_factor}\NormalTok{(Keywords)) }\SpecialCharTok{\%\textgreater{}\%}
  \FunctionTok{mutate}\NormalTok{(}\AttributeTok{Keywords =} \FunctionTok{fct\_explicit\_na}\NormalTok{(Keywords, }\StringTok{"wo\_tag"}\NormalTok{)) }\SpecialCharTok{\%\textgreater{}\%} \CommentTok{\# when pix has no tag}
  \FunctionTok{select}\NormalTok{(SourceFile, FileName, Keywords) }\SpecialCharTok{\%\textgreater{}\%}
  \FunctionTok{mutate}\NormalTok{(}\AttributeTok{Keywords =} \FunctionTok{fct\_recode}\NormalTok{(Keywords,}
                               \StringTok{"chat"} \OtherTok{=} \StringTok{"chat forestier"}\NormalTok{,}
                               \StringTok{"lievre"} \OtherTok{=} \StringTok{"lièvre"}\NormalTok{,}
                               \StringTok{"vehicule"} \OtherTok{=} \StringTok{"véhicule"}\NormalTok{,}
                               \StringTok{"ni"} \OtherTok{=} \StringTok{"Non identifié"}\NormalTok{)) }\SpecialCharTok{\%\textgreater{}\%}
  \FunctionTok{filter}\NormalTok{(}\SpecialCharTok{!}\NormalTok{(Keywords }\SpecialCharTok{\%in\%} \FunctionTok{c}\NormalTok{(}\StringTok{"ni"}\NormalTok{, }\StringTok{"wo\_tag"}\NormalTok{)))}
\end{Highlighting}
\end{Shaded}

\begin{table}

\caption{\label{tab:unnamed-chunk-4}Species considered, and number of images with these species in them.}
\centering
\begin{tabular}[t]{l|r}
\hline
Keywords & n\\
\hline
humain & 143\\
\hline
vehicule & 135\\
\hline
renard & 58\\
\hline
sangliers & 33\\
\hline
chasseur & 17\\
\hline
chien & 14\\
\hline
lynx & 13\\
\hline
chevreuil & 13\\
\hline
chamois & 12\\
\hline
blaireaux & 10\\
\hline
chat & 8\\
\hline
lievre & 4\\
\hline
fouine & 1\\
\hline
cavalier & 1\\
\hline
\end{tabular}
\end{table}

Then we pick 80\(\%\) of the images for training in each category, the
rest being used for validation.

\begin{Shaded}
\begin{Highlighting}[]
\CommentTok{\# training dataset}
\NormalTok{pix\_train }\OtherTok{\textless{}{-}}\NormalTok{ labels }\SpecialCharTok{\%\textgreater{}\%}
  \FunctionTok{select}\NormalTok{(SourceFile, FileName, Keywords) }\SpecialCharTok{\%\textgreater{}\%}
  \FunctionTok{group\_by}\NormalTok{(Keywords) }\SpecialCharTok{\%\textgreater{}\%}
  \FunctionTok{filter}\NormalTok{(}\FunctionTok{between}\NormalTok{(}\FunctionTok{row\_number}\NormalTok{(), }\DecValTok{1}\NormalTok{, }\FunctionTok{floor}\NormalTok{(}\FunctionTok{n}\NormalTok{()}\SpecialCharTok{*}\DecValTok{80}\SpecialCharTok{/}\DecValTok{100}\NormalTok{))) }\CommentTok{\# 80\% per category}
\CommentTok{\# validation dataset}
\NormalTok{pix\_valid }\OtherTok{\textless{}{-}}\NormalTok{ labels }\SpecialCharTok{\%\textgreater{}\%}
  \FunctionTok{group\_by}\NormalTok{(Keywords) }\SpecialCharTok{\%\textgreater{}\%}
  \FunctionTok{filter}\NormalTok{(}\FunctionTok{between}\NormalTok{(}\FunctionTok{row\_number}\NormalTok{(), }\FunctionTok{floor}\NormalTok{(}\FunctionTok{n}\NormalTok{()}\SpecialCharTok{*}\DecValTok{80}\SpecialCharTok{/}\DecValTok{100}\NormalTok{) }\SpecialCharTok{+} \DecValTok{1}\NormalTok{, }\FunctionTok{n}\NormalTok{()))}
\end{Highlighting}
\end{Shaded}

Eventually, we store these images in two distinct directories named
\texttt{train} and \texttt{valid}.

\begin{Shaded}
\begin{Highlighting}[]
\CommentTok{\# create dir train/ and copy pix there, organised by categories}
\FunctionTok{dir.create}\NormalTok{(}\StringTok{\textquotesingle{}pix/train\textquotesingle{}}\NormalTok{) }\CommentTok{\# create training directory}
\ControlFlowTok{for}\NormalTok{ (i }\ControlFlowTok{in} \FunctionTok{levels}\NormalTok{(}\FunctionTok{fct\_drop}\NormalTok{(pix\_train}\SpecialCharTok{$}\NormalTok{Keywords))) }\FunctionTok{dir.create}\NormalTok{(}\FunctionTok{paste0}\NormalTok{(}\StringTok{\textquotesingle{}pix/train/\textquotesingle{}}\NormalTok{,i)) }\CommentTok{\# create dir for labels}
\ControlFlowTok{for}\NormalTok{ (i }\ControlFlowTok{in} \DecValTok{1}\SpecialCharTok{:}\FunctionTok{nrow}\NormalTok{(pix\_train))\{}
    \FunctionTok{file.copy}\NormalTok{(}\FunctionTok{as.character}\NormalTok{(pix\_train}\SpecialCharTok{$}\NormalTok{SourceFile[i]),}
              \FunctionTok{paste0}\NormalTok{(}\StringTok{\textquotesingle{}pix/train/\textquotesingle{}}\NormalTok{, }\FunctionTok{as.character}\NormalTok{(pix\_train}\SpecialCharTok{$}\NormalTok{Keywords[i]))) }\CommentTok{\# copy pix in corresp dir}
\NormalTok{\}}
\CommentTok{\# create dir valid/ and copy pix there, organised by categories.}
\FunctionTok{dir.create}\NormalTok{(}\StringTok{\textquotesingle{}pix/valid\textquotesingle{}}\NormalTok{) }\CommentTok{\# create validation dir}
\ControlFlowTok{for}\NormalTok{ (i }\ControlFlowTok{in} \FunctionTok{levels}\NormalTok{(}\FunctionTok{fct\_drop}\NormalTok{(pix\_train}\SpecialCharTok{$}\NormalTok{Keywords))) }\FunctionTok{dir.create}\NormalTok{(}\FunctionTok{paste0}\NormalTok{(}\StringTok{\textquotesingle{}pix/valid/\textquotesingle{}}\NormalTok{,i)) }\CommentTok{\# create dir for labels}
\ControlFlowTok{for}\NormalTok{ (i }\ControlFlowTok{in} \DecValTok{1}\SpecialCharTok{:}\FunctionTok{nrow}\NormalTok{(pix\_valid))\{}
    \FunctionTok{file.copy}\NormalTok{(}\FunctionTok{as.character}\NormalTok{(pix\_valid}\SpecialCharTok{$}\NormalTok{SourceFile[i]),}
              \FunctionTok{paste0}\NormalTok{(}\StringTok{\textquotesingle{}pix/valid/\textquotesingle{}}\NormalTok{, }\FunctionTok{as.character}\NormalTok{(pix\_valid}\SpecialCharTok{$}\NormalTok{Keywords[i]))) }\CommentTok{\# copy pix in corresp dir}
\NormalTok{\}}
\CommentTok{\# delete pictures in valid/ directory for which we did not train the model}
\NormalTok{to\_be\_deleted }\OtherTok{\textless{}{-}} \FunctionTok{setdiff}\NormalTok{(}\FunctionTok{levels}\NormalTok{(}\FunctionTok{fct\_drop}\NormalTok{(pix\_valid}\SpecialCharTok{$}\NormalTok{Keywords)), }\FunctionTok{levels}\NormalTok{(}\FunctionTok{fct\_drop}\NormalTok{(pix\_train}\SpecialCharTok{$}\NormalTok{Keywords)))}
\ControlFlowTok{if}\NormalTok{ (}\SpecialCharTok{!}\FunctionTok{is\_empty}\NormalTok{(to\_be\_deleted)) \{}
  \ControlFlowTok{for}\NormalTok{ (i }\ControlFlowTok{in} \DecValTok{1}\SpecialCharTok{:}\FunctionTok{length}\NormalTok{(to\_be\_deleted))\{}
    \FunctionTok{unlink}\NormalTok{(}\FunctionTok{paste0}\NormalTok{(}\StringTok{\textquotesingle{}pix/valid/\textquotesingle{}}\NormalTok{, to\_be\_deleted[i]))}
\NormalTok{  \}}
\NormalTok{\}}
\end{Highlighting}
\end{Shaded}

What is the sample size of these two datasets?

\begin{Shaded}
\begin{Highlighting}[]
\FunctionTok{bind\_rows}\NormalTok{(}\StringTok{"training"} \OtherTok{=}\NormalTok{ pix\_train, }\StringTok{"validation"} \OtherTok{=}\NormalTok{ pix\_valid, }\AttributeTok{.id =} \StringTok{"dataset"}\NormalTok{) }\SpecialCharTok{\%\textgreater{}\%}
  \FunctionTok{group\_by}\NormalTok{(dataset) }\SpecialCharTok{\%\textgreater{}\%}
  \FunctionTok{count}\NormalTok{(Keywords) }\SpecialCharTok{\%\textgreater{}\%}
  \FunctionTok{rename}\NormalTok{(}\AttributeTok{category =}\NormalTok{ Keywords) }\SpecialCharTok{\%\textgreater{}\%}
  \FunctionTok{kable}\NormalTok{(}\AttributeTok{caption =} \StringTok{"Sample size (n) for the training and validation datasets."}\NormalTok{) }\SpecialCharTok{\%\textgreater{}\%}
  \FunctionTok{kable\_styling}\NormalTok{()}
\end{Highlighting}
\end{Shaded}

\begin{table}

\caption{\label{tab:unnamed-chunk-7}Sample size (n) for the training and validation datasets.}
\centering
\begin{tabular}[t]{l|l|r}
\hline
dataset & category & n\\
\hline
training & humain & 114\\
\hline
training & vehicule & 108\\
\hline
training & chamois & 9\\
\hline
training & blaireaux & 8\\
\hline
training & sangliers & 26\\
\hline
training & renard & 46\\
\hline
training & chasseur & 13\\
\hline
training & lynx & 10\\
\hline
training & chien & 11\\
\hline
training & chat & 6\\
\hline
training & chevreuil & 10\\
\hline
training & lievre & 3\\
\hline
validation & humain & 29\\
\hline
validation & vehicule & 27\\
\hline
validation & chamois & 3\\
\hline
validation & blaireaux & 2\\
\hline
validation & sangliers & 7\\
\hline
validation & renard & 12\\
\hline
validation & chasseur & 4\\
\hline
validation & lynx & 3\\
\hline
validation & chien & 3\\
\hline
validation & fouine & 1\\
\hline
validation & chat & 2\\
\hline
validation & chevreuil & 3\\
\hline
validation & lievre & 1\\
\hline
validation & cavalier & 1\\
\hline
\end{tabular}
\end{table}

\hypertarget{transfer-learning}{%
\subsection{Transfer learning}\label{transfer-learning}}

We proceed with transfer learning using images from the Jura county (or
a subsample more exactly). We first load images and apply standard
transformations to improve training (flip, rotate, zoom, rotate, ligth
transform).

\begin{Shaded}
\begin{Highlighting}[]
\NormalTok{dls }\OtherTok{\textless{}{-}} \FunctionTok{ImageDataLoaders\_from\_folder}\NormalTok{(}
  \AttributeTok{path =} \StringTok{"pix/"}\NormalTok{,}
  \AttributeTok{train =} \StringTok{"train"}\NormalTok{,}
  \AttributeTok{valid =} \StringTok{"valid"}\NormalTok{,}
  \AttributeTok{item\_tfms =} \FunctionTok{Resize}\NormalTok{(}\AttributeTok{size =} \DecValTok{460}\NormalTok{),}
  \AttributeTok{bs =} \DecValTok{10}\NormalTok{,}
  \AttributeTok{batch\_tfms =} \FunctionTok{list}\NormalTok{(}\FunctionTok{aug\_transforms}\NormalTok{(}\AttributeTok{size =} \DecValTok{224}\NormalTok{,}
                                   \AttributeTok{min\_scale =} \FloatTok{0.75}\NormalTok{), }\CommentTok{\# transformation}
                    \FunctionTok{Normalize\_from\_stats}\NormalTok{( }\FunctionTok{imagenet\_stats}\NormalTok{() )),}
  \AttributeTok{num\_workers =} \DecValTok{0}\NormalTok{,}
  \AttributeTok{ImageFile.LOAD\_TRUNCATED\_IMAGES =} \ConstantTok{TRUE}\NormalTok{)}
\end{Highlighting}
\end{Shaded}

Then we get the model architecture. For the sake of illustration, we use
a resnet18 here, but we used a resnet50 to get the full results
presented in the main text.

\begin{Shaded}
\begin{Highlighting}[]
\NormalTok{learn }\OtherTok{\textless{}{-}} \FunctionTok{cnn\_learner}\NormalTok{(}\AttributeTok{dls =}\NormalTok{ dls,}
                     \AttributeTok{arch =} \FunctionTok{resnet18}\NormalTok{(),}
                     \AttributeTok{metrics =} \FunctionTok{list}\NormalTok{(accuracy, error\_rate))}
\end{Highlighting}
\end{Shaded}

Now we are ready to train our model. Again, for the sake of
illustration, we use only 2 epochs here, but used 20 epochs to get the
full results presented in the main text. With all pictures and a
resnet50, it took 75 minutes per epoch approximatively on a Mac with a
2.4Ghz processor and 64Go memory, and less than half an hour on a
machine with GPU. On this reduced dataset, it took a bit more than a
minute per epoch on the same Mac. Note that we save the model after each
epoch for later use.

\begin{Shaded}
\begin{Highlighting}[]
\NormalTok{one\_cycle }\OtherTok{\textless{}{-}}\NormalTok{ learn }\SpecialCharTok{\%\textgreater{}\%}
  \FunctionTok{fit\_one\_cycle}\NormalTok{(}\DecValTok{2}\NormalTok{, }\AttributeTok{cbs =} \FunctionTok{SaveModelCallback}\NormalTok{(}\AttributeTok{every\_epoch =} \ConstantTok{TRUE}\NormalTok{,}
                                           \AttributeTok{fname =} \StringTok{\textquotesingle{}model\textquotesingle{}}\NormalTok{))}
\end{Highlighting}
\end{Shaded}

\begin{verbatim}
## epoch   train_loss   valid_loss   accuracy   error_rate   time 
## ------  -----------  -----------  ---------  -----------  -----
## 0       2.447199     0.850911     0.760417   0.239583     01:24 
## 1       1.736201     0.666055     0.781250   0.218750     01:25
\end{verbatim}

\begin{Shaded}
\begin{Highlighting}[]
\NormalTok{one\_cycle}
\end{Highlighting}
\end{Shaded}

\begin{verbatim}
##   epoch train_loss valid_loss  accuracy error_rate
## 1     0   2.447199  0.8509111 0.7604167  0.2395833
## 2     1   1.736201  0.6660554 0.7812500  0.2187500
\end{verbatim}

We may dig a bit deeper in training performances by loading the best
model, here \texttt{model\_1.pth}, and display some metrics for each
species.

\begin{Shaded}
\begin{Highlighting}[]
\NormalTok{learn}\SpecialCharTok{$}\FunctionTok{load}\NormalTok{(}\StringTok{"model\_1"}\NormalTok{)}
\end{Highlighting}
\end{Shaded}

\begin{verbatim}
## Sequential(
##   (0): Sequential(
##     (0): Conv2d(3, 64, kernel_size=(7, 7), stride=(2, 2), padding=(3, 3), bias=False)
##     (1): BatchNorm2d(64, eps=1e-05, momentum=0.1, affine=True, track_running_stats=True)
##     (2): ReLU(inplace=True)
##     (3): MaxPool2d(kernel_size=3, stride=2, padding=1, dilation=1, ceil_mode=False)
##     (4): Sequential(
##       (0): BasicBlock(
##         (conv1): Conv2d(64, 64, kernel_size=(3, 3), stride=(1, 1), padding=(1, 1), bias=False)
##         (bn1): BatchNorm2d(64, eps=1e-05, momentum=0.1, affine=True, track_running_stats=True)
##         (relu): ReLU(inplace=True)
##         (conv2): Conv2d(64, 64, kernel_size=(3, 3), stride=(1, 1), padding=(1, 1), bias=False)
##         (bn2): BatchNorm2d(64, eps=1e-05, momentum=0.1, affine=True, track_running_stats=True)
##       )
##       (1): BasicBlock(
##         (conv1): Conv2d(64, 64, kernel_size=(3, 3), stride=(1, 1), padding=(1, 1), bias=False)
##         (bn1): BatchNorm2d(64, eps=1e-05, momentum=0.1, affine=True, track_running_stats=True)
##         (relu): ReLU(inplace=True)
##         (conv2): Conv2d(64, 64, kernel_size=(3, 3), stride=(1, 1), padding=(1, 1), bias=False)
##         (bn2): BatchNorm2d(64, eps=1e-05, momentum=0.1, affine=True, track_running_stats=True)
##       )
##     )
##     (5): Sequential(
##       (0): BasicBlock(
##         (conv1): Conv2d(64, 128, kernel_size=(3, 3), stride=(2, 2), padding=(1, 1), bias=False)
##         (bn1): BatchNorm2d(128, eps=1e-05, momentum=0.1, affine=True, track_running_stats=True)
##         (relu): ReLU(inplace=True)
##         (conv2): Conv2d(128, 128, kernel_size=(3, 3), stride=(1, 1), padding=(1, 1), bias=False)
##         (bn2): BatchNorm2d(128, eps=1e-05, momentum=0.1, affine=True, track_running_stats=True)
##         (downsample): Sequential(
##           (0): Conv2d(64, 128, kernel_size=(1, 1), stride=(2, 2), bias=False)
##           (1): BatchNorm2d(128, eps=1e-05, momentum=0.1, affine=True, track_running_stats=True)
##         )
##       )
##       (1): BasicBlock(
##         (conv1): Conv2d(128, 128, kernel_size=(3, 3), stride=(1, 1), padding=(1, 1), bias=False)
##         (bn1): BatchNorm2d(128, eps=1e-05, momentum=0.1, affine=True, track_running_stats=True)
##         (relu): ReLU(inplace=True)
##         (conv2): Conv2d(128, 128, kernel_size=(3, 3), stride=(1, 1), padding=(1, 1), bias=False)
##         (bn2): BatchNorm2d(128, eps=1e-05, momentum=0.1, affine=True, track_running_stats=True)
##       )
##     )
##     (6): Sequential(
##       (0): BasicBlock(
##         (conv1): Conv2d(128, 256, kernel_size=(3, 3), stride=(2, 2), padding=(1, 1), bias=False)
##         (bn1): BatchNorm2d(256, eps=1e-05, momentum=0.1, affine=True, track_running_stats=True)
##         (relu): ReLU(inplace=True)
##         (conv2): Conv2d(256, 256, kernel_size=(3, 3), stride=(1, 1), padding=(1, 1), bias=False)
##         (bn2): BatchNorm2d(256, eps=1e-05, momentum=0.1, affine=True, track_running_stats=True)
##         (downsample): Sequential(
##           (0): Conv2d(128, 256, kernel_size=(1, 1), stride=(2, 2), bias=False)
##           (1): BatchNorm2d(256, eps=1e-05, momentum=0.1, affine=True, track_running_stats=True)
##         )
##       )
##       (1): BasicBlock(
##         (conv1): Conv2d(256, 256, kernel_size=(3, 3), stride=(1, 1), padding=(1, 1), bias=False)
##         (bn1): BatchNorm2d(256, eps=1e-05, momentum=0.1, affine=True, track_running_stats=True)
##         (relu): ReLU(inplace=True)
##         (conv2): Conv2d(256, 256, kernel_size=(3, 3), stride=(1, 1), padding=(1, 1), bias=False)
##         (bn2): BatchNorm2d(256, eps=1e-05, momentum=0.1, affine=True, track_running_stats=True)
##       )
##     )
##     (7): Sequential(
##       (0): BasicBlock(
##         (conv1): Conv2d(256, 512, kernel_size=(3, 3), stride=(2, 2), padding=(1, 1), bias=False)
##         (bn1): BatchNorm2d(512, eps=1e-05, momentum=0.1, affine=True, track_running_stats=True)
##         (relu): ReLU(inplace=True)
##         (conv2): Conv2d(512, 512, kernel_size=(3, 3), stride=(1, 1), padding=(1, 1), bias=False)
##         (bn2): BatchNorm2d(512, eps=1e-05, momentum=0.1, affine=True, track_running_stats=True)
##         (downsample): Sequential(
##           (0): Conv2d(256, 512, kernel_size=(1, 1), stride=(2, 2), bias=False)
##           (1): BatchNorm2d(512, eps=1e-05, momentum=0.1, affine=True, track_running_stats=True)
##         )
##       )
##       (1): BasicBlock(
##         (conv1): Conv2d(512, 512, kernel_size=(3, 3), stride=(1, 1), padding=(1, 1), bias=False)
##         (bn1): BatchNorm2d(512, eps=1e-05, momentum=0.1, affine=True, track_running_stats=True)
##         (relu): ReLU(inplace=True)
##         (conv2): Conv2d(512, 512, kernel_size=(3, 3), stride=(1, 1), padding=(1, 1), bias=False)
##         (bn2): BatchNorm2d(512, eps=1e-05, momentum=0.1, affine=True, track_running_stats=True)
##       )
##     )
##   )
##   (1): Sequential(
##     (0): AdaptiveConcatPool2d(
##       (ap): AdaptiveAvgPool2d(output_size=1)
##       (mp): AdaptiveMaxPool2d(output_size=1)
##     )
##     (1): Flatten(full=False)
##     (2): BatchNorm1d(1024, eps=1e-05, momentum=0.1, affine=True, track_running_stats=True)
##     (3): Dropout(p=0.25, inplace=False)
##     (4): Linear(in_features=1024, out_features=512, bias=False)
##     (5): ReLU(inplace=True)
##     (6): BatchNorm1d(512, eps=1e-05, momentum=0.1, affine=True, track_running_stats=True)
##     (7): Dropout(p=0.5, inplace=False)
##     (8): Linear(in_features=512, out_features=12, bias=False)
##   )
## )
\end{verbatim}

\begin{Shaded}
\begin{Highlighting}[]
\NormalTok{interp }\OtherTok{\textless{}{-}} \FunctionTok{ClassificationInterpretation\_from\_learner}\NormalTok{(learn)}
\NormalTok{interp}\SpecialCharTok{$}\FunctionTok{print\_classification\_report}\NormalTok{()}
\end{Highlighting}
\end{Shaded}

We may extract the categories that get the most confused.

\begin{Shaded}
\begin{Highlighting}[]
\NormalTok{interp }\SpecialCharTok{\%\textgreater{}\%} \FunctionTok{most\_confused}\NormalTok{()}
\end{Highlighting}
\end{Shaded}

\begin{verbatim}
##           V1        V2 V3
## 1     humain  vehicule  4
## 2   chasseur    humain  3
## 3  blaireaux    renard  1
## 4  blaireaux sangliers  1
## 5       chat    renard  1
## 6       chat sangliers  1
## 7  chevreuil     chien  1
## 8  chevreuil    renard  1
## 9      chien   chamois  1
## 10     chien sangliers  1
## 11    humain  chasseur  1
## 12    lievre    renard  1
## 13    renard blaireaux  1
## 14    renard sangliers  1
## 15 sangliers    renard  1
## 16  vehicule    humain  1
\end{verbatim}

\hypertarget{transferability}{%
\subsection{Transferability}\label{transferability}}

In this section, we show how to use our freshly trained model to label
images that were taken in another study site in the Ain county, and not
used to train our model. First, we get the path to the images.

\begin{Shaded}
\begin{Highlighting}[]
\NormalTok{fls }\OtherTok{\textless{}{-}} \FunctionTok{list.files}\NormalTok{(}\AttributeTok{path =} \StringTok{"pix/pixAin"}\NormalTok{,}
                  \AttributeTok{full.names =} \ConstantTok{TRUE}\NormalTok{,}
                  \AttributeTok{recursive =} \ConstantTok{TRUE}\NormalTok{)}
\end{Highlighting}
\end{Shaded}

Then we carry out prediction, and compare to the truth.

\begin{Shaded}
\begin{Highlighting}[]
\NormalTok{predicted }\OtherTok{\textless{}{-}} \FunctionTok{character}\NormalTok{(}\DecValTok{3}\NormalTok{)}
\NormalTok{categories }\OtherTok{\textless{}{-}}\NormalTok{ interp}\SpecialCharTok{$}\NormalTok{vocab }\SpecialCharTok{\%\textgreater{}\%}
  \FunctionTok{str\_replace\_all}\NormalTok{(}\StringTok{"[[:punct:]]"}\NormalTok{, }\StringTok{" "}\NormalTok{) }\SpecialCharTok{\%\textgreater{}\%}
  \FunctionTok{str\_trim}\NormalTok{() }\SpecialCharTok{\%\textgreater{}\%}
  \FunctionTok{str\_split}\NormalTok{(}\StringTok{"   "}\NormalTok{) }\SpecialCharTok{\%\textgreater{}\%}
  \FunctionTok{unlist}\NormalTok{()}
\ControlFlowTok{for}\NormalTok{ (i }\ControlFlowTok{in} \DecValTok{1}\SpecialCharTok{:}\FunctionTok{length}\NormalTok{(fls))\{}
\NormalTok{  result }\OtherTok{\textless{}{-}}\NormalTok{ learn }\SpecialCharTok{\%\textgreater{}\%} \FunctionTok{predict}\NormalTok{(fls[i]) }\CommentTok{\# make prediction}
\NormalTok{  result[[}\DecValTok{3}\NormalTok{]] }\SpecialCharTok{\%\textgreater{}\%}
    \FunctionTok{str\_extract}\NormalTok{(}\StringTok{"}\SpecialCharTok{\textbackslash{}\textbackslash{}}\StringTok{d+"}\NormalTok{) }\SpecialCharTok{\%\textgreater{}\%}
    \FunctionTok{as.integer}\NormalTok{() }\OtherTok{{-}\textgreater{}}\NormalTok{ index }\CommentTok{\# extract relevant info}
\NormalTok{  predicted[i] }\OtherTok{\textless{}{-}}\NormalTok{ categories[index }\SpecialCharTok{+} \DecValTok{1}\NormalTok{] }\CommentTok{\# match it with categories}
\NormalTok{\}}
\FunctionTok{data.frame}\NormalTok{(}\AttributeTok{truth =} \FunctionTok{c}\NormalTok{(}\StringTok{"lynx"}\NormalTok{, }\StringTok{"roe deer"}\NormalTok{, }\StringTok{"wild boar"}\NormalTok{),}
           \AttributeTok{prediction =}\NormalTok{ predicted) }\SpecialCharTok{\%\textgreater{}\%}
  \FunctionTok{kable}\NormalTok{(}\AttributeTok{caption =} \StringTok{"Comparison of the predictions vs. ground truth."}\NormalTok{) }\SpecialCharTok{\%\textgreater{}\%}
  \FunctionTok{kable\_styling}\NormalTok{()}
\end{Highlighting}
\end{Shaded}

\begin{table}

\caption{\label{tab:unnamed-chunk-14}Comparison of the predictions vs. ground truth.}
\centering
\begin{tabular}[t]{l|l}
\hline
truth & prediction\\
\hline
lynx & chevreuil\\
\hline
roe deer & chamois\\
\hline
wild boar & sangliers\\
\hline
\end{tabular}
\end{table}

\hypertarget{session-information}{%
\section{Session information}\label{session-information}}

\begin{verbatim}
## R version 4.1.0 (2021-05-18)
## Platform: x86_64-apple-darwin17.0 (64-bit)
## Running under: macOS Catalina 10.15.7
## 
## Matrix products: default
## BLAS:   /Library/Frameworks/R.framework/Versions/4.1/Resources/lib/libRblas.dylib
## LAPACK: /Library/Frameworks/R.framework/Versions/4.1/Resources/lib/libRlapack.dylib
## 
## locale:
## [1] fr_FR.UTF-8/fr_FR.UTF-8/fr_FR.UTF-8/C/fr_FR.UTF-8/fr_FR.UTF-8
## 
## attached base packages:
## [1] stats     graphics  grDevices utils     datasets  methods   base     
## 
## other attached packages:
##  [1] exifr_0.3.2         unmarked_1.1.1.9006 lattice_0.20-44    
##  [4] janitor_2.1.0       highcharter_0.8.2   fastai_2.0.9       
##  [7] ggtext_0.1.1        wesanderson_0.3.6   kableExtra_1.3.4   
## [10] stringi_1.7.3       lubridate_1.7.10    cowplot_1.1.1      
## [13] sf_1.0-2            forcats_0.5.1       stringr_1.4.0      
## [16] dplyr_1.0.7         purrr_0.3.4.9000    readr_2.0.0        
## [19] tidyr_1.1.3         tibble_3.1.3        ggplot2_3.3.5      
## [22] tidyverse_1.3.1    
## 
## loaded via a namespace (and not attached):
##   [1] minqa_1.2.4        colorspace_2.0-2   ggsignif_0.6.2    
##   [4] rio_0.5.27         ellipsis_0.3.2     class_7.3-19      
##   [7] snakecase_0.11.0   markdown_1.1       fs_1.5.0          
##  [10] gridtext_0.1.4     rstudioapi_0.13    proxy_0.4-26      
##  [13] ggpubr_0.4.0       farver_2.1.0       bit64_4.0.5       
##  [16] fansi_0.5.0        xml2_1.3.2         codetools_0.2-18  
##  [19] splines_4.1.0      knitr_1.33         rlist_0.4.6.1     
##  [22] jsonlite_1.7.2     nloptr_1.2.2.2     broom_0.7.9       
##  [25] dbplyr_2.1.1       png_0.1-7          compiler_4.1.0    
##  [28] httr_1.4.2         backports_1.2.1    assertthat_0.2.1  
##  [31] Matrix_1.3-3       cli_3.0.1          htmltools_0.5.1.1 
##  [34] tools_4.1.0        igraph_1.2.6       gtable_0.3.0      
##  [37] glue_1.4.2         rappdirs_0.3.3     Rcpp_1.0.7        
##  [40] carData_3.0-4      cellranger_1.1.0   raster_3.4-13     
##  [43] vctrs_0.3.8        nlme_3.1-152       svglite_2.0.0     
##  [46] xfun_0.25          ps_1.6.0           openxlsx_4.2.4    
##  [49] lme4_1.1-27.1      rvest_1.0.1        lifecycle_1.0.0   
##  [52] rstatix_0.7.0      MASS_7.3-54        zoo_1.8-9         
##  [55] scales_1.1.1       vroom_1.5.4        ragg_1.1.3        
##  [58] hms_1.1.0          parallel_4.1.0     yaml_2.2.1        
##  [61] quantmod_0.4.18    curl_4.3.2         reticulate_1.20   
##  [64] pbapply_1.4-3      highr_0.9          e1071_1.7-8       
##  [67] TTR_0.24.2         zip_2.2.0          boot_1.3-28       
##  [70] rlang_0.4.11       pkgconfig_2.0.3    systemfonts_1.0.2 
##  [73] evaluate_0.14      labeling_0.4.2     htmlwidgets_1.5.3 
##  [76] bit_4.0.4          tidyselect_1.1.1   processx_3.5.2    
##  [79] plyr_1.8.6         magrittr_2.0.1     R6_2.5.0          
##  [82] generics_0.1.0     DBI_1.1.1          foreign_0.8-81    
##  [85] pillar_1.6.2       haven_2.4.3        withr_2.4.2       
##  [88] units_0.7-2        xts_0.12.1         abind_1.4-5       
##  [91] sp_1.4-5           car_3.0-11         modelr_0.1.8      
##  [94] crayon_1.4.1       KernSmooth_2.23-20 utf8_1.2.2        
##  [97] tzdb_0.1.2         rmarkdown_2.10     jpeg_0.1-9        
## [100] grid_4.1.0         readxl_1.3.1       data.table_1.14.0 
## [103] callr_3.7.0        reprex_2.0.1       digest_0.6.27     
## [106] classInt_0.4-3     webshot_0.5.2      textshaping_0.3.5 
## [109] munsell_0.5.0      viridisLite_0.4.0
\end{verbatim}

\hypertarget{acknowledgments}{%
\section{Acknowledgments}\label{acknowledgments}}

We warmly thank Mathieu Massaviol, Remy Dernat and Khalid Belkhir for
their help in using GPU machines on the Montpellier Bioinformatics
Biodiversity platform, Julien Renoult for helpful discussions, Delphine
Dinouart and Chloé Quillard for their precious help in manually tagging
the images, and Vincent Miele for having inspired this work, and his
help and support along the way. We also thank the staff of the
Federations of Hunters from the Jura and Ain counties, hunters who
helped to find locations for camera traps and volunteers who contributed
in collecting data. Last, we thank Auvergne-Rhône-Alpes Region, Ain and
Jura departmental Councils, The French National Federation of Hunters,
French Environmental Ministry based in Auvergne-Rhone-Alpes and
Bourgogne Franche-Comté Region and the French Office for Biodiversity
for funding the Lynx Predator Prey Program. This work was also partly
funded by the French National Research Agency (grant ANR-16-CE02-0007).

\hypertarget{references}{%
\section*{References}\label{references}}
\addcontentsline{toc}{section}{References}

\hypertarget{refs}{}
\begin{CSLReferences}{1}{0}
\leavevmode\hypertarget{ref-reticulate_ref}{}%
Allaire, JJ, Kevin Ushey, Yuan Tang, and Dirk Eddelbuettel. 2017.
\emph{Reticulate: R Interface to Python}.
\url{https://github.com/rstudio/reticulate}.

\leavevmode\hypertarget{ref-baraniuk_science_2020}{}%
Baraniuk, Richard, David Donoho, and Matan Gavish. 2020. {``The Science
of Deep Learning.''} \emph{Proceedings of the National Academy of
Sciences} 117 (48): 30029--32.
\url{https://doi.org/10.1073/pnas.2020596117}.

\leavevmode\hypertarget{ref-beery_recognition_2018}{}%
Beery, Sara, Grant van Horn, and Pietro Perona. 2018. {``Recognition in
{Terra} {Incognita}.''} \emph{arXiv:1807.04975 {[}Cs, q-Bio{]}}, July.
\url{http://arxiv.org/abs/1807.04975}.

\leavevmode\hypertarget{ref-breitenmoser_large_1998}{}%
Breitenmoser, Urs. 1998. {``Large Predators in the {Alps}: {The} Fall
and Rise of Man's Competitors.''} \emph{Biological Conservation},
Conservation {Biology} and {Biodiversity} {Strategies}, 83 (3): 279--89.
\url{https://doi.org/10.1016/S0006-3207(97)00084-0}.

\leavevmode\hypertarget{ref-chambert2018}{}%
Chambert, Thierry, Evan H. Campbell Grant, David A. W. Miller, James D.
Nichols, Kevin P. Mulder, and Adrianne B. Brand. 2018. {``Two-Species
Occupancy Modelling Accounting for Species Misidentification and
Non-Detection.''} \emph{Methods in Ecology and Evolution} 9 (6):
1468--77. https://doi.org/\url{https://doi.org/10.1111/2041-210X.12985}.

\leavevmode\hypertarget{ref-christin_applications_2019}{}%
Christin, Sylvain, Éric Hervet, and Nicolas Lecomte. 2019.
{``Applications for Deep Learning in Ecology.''} Edited by Hao Ye.
\emph{Methods in Ecology and Evolution} 10 (10): 1632--44.
\url{https://doi.org/10.1111/2041-210X.13256}.

\leavevmode\hypertarget{ref-clipp2021}{}%
Clipp, Hannah L., Amber L. Evans, Brin E. Kessinger, K. Kellner, and
Christopher T. Rota. 2021. {``A Penalized Likelihood for Multi-Species
Occupancy Models Improves Predictions of Species Interactions.''}
\emph{Ecology}.

\leavevmode\hypertarget{ref-dai_multivariate_2013}{}%
Dai, Bin, Shilin Ding, and Grace Wahba. 2013. {``Multivariate
{Bernoulli} Distribution.''} \emph{Bernoulli} 19 (4).
\url{https://doi.org/10.3150/12-BEJSP10}.

\leavevmode\hypertarget{ref-Duggan2021}{}%
Duggan, Matthew T., Melissa F. Groleau, Ethan P. Shealy, Lillian S.
Self, Taylor E. Utter, Matthew M. Waller, Bryan C. Hall, Chris G. Stone,
Layne L. Anderson, and Timothy A. Mousseau. 2021. {``An Approach to
Rapid Processing of Camera Trap Images with Minimal Human Input.''}
\emph{Ecology and Evolution}.
https://doi.org/\url{https://doi.org/10.1002/ece3.7970}.

\leavevmode\hypertarget{ref-unmarkedFiske}{}%
Fiske, Ian, and Richard Chandler. 2011. {``{unmarked}: An {R} Package
for Fitting Hierarchical Models of Wildlife Occurrence and Abundance.''}
\emph{Journal of Statistical Software} 43 (10): 1--23.
\url{https://www.jstatsoft.org/v43/i10/}.

\leavevmode\hypertarget{ref-gimenez_statistical_2014}{}%
Gimenez, Olivier, Stephen T. Buckland, Byron J. T. Morgan, Nicolas Bez,
Sophie Bertrand, Rémi Choquet, Stéphane Dray, et al. 2014.
{``Statistical Ecology Comes of Age.''} \emph{Biology Letters} 10 (12):
20140698. \url{https://doi.org/10.1098/rsbl.2014.0698}.

\leavevmode\hypertarget{ref-he_deep_2016}{}%
He, Kaiming, Xiangyu Zhang, Shaoqing Ren, and Jian Sun. 2016. {``Deep
{Residual} {Learning} for {Image} {Recognition}.''} In \emph{2016 {IEEE}
{Conference} on {Computer} {Vision} and {Pattern} {Recognition}
({CVPR})}, 770--78. \url{https://doi.org/10.1109/CVPR.2016.90}.

\leavevmode\hypertarget{ref-NIPS2012_4824}{}%
Krizhevsky, Alex, Ilya Sutskever, and Geoffrey E. Hinton. 2012.
{``ImageNet Classification with Deep Convolutional Neural Networks.''}
In \emph{Advances in Neural Information Processing Systems 25}, edited
by F. Pereira, C. J. C. Burges, L. Bottou, and K. Q. Weinberger,
1097--1105. Curran Associates, Inc.

\leavevmode\hypertarget{ref-lahoz-monfort_comprehensive_2021}{}%
Lahoz-Monfort, José J, and Michael J L Magrath. 2021. {``A
{Comprehensive} {Overview} of {Technologies} for {Species} and {Habitat}
{Monitoring} and {Conservation}.''} \emph{BioScience}.
\url{https://doi.org/10.1093/biosci/biab073}.

\leavevmode\hypertarget{ref-lai_evaluating_2019}{}%
Lai, Jiangshan, Christopher J. Lortie, Robert A. Muenchen, Jian Yang,
and Keping Ma. 2019. {``Evaluating the Popularity of {R} in Ecology.''}
\emph{Ecosphere} 10 (1). \url{https://doi.org/10.1002/ecs2.2567}.

\leavevmode\hypertarget{ref-lamba_deep_2019}{}%
Lamba, Aakash, Phillip Cassey, Ramesh Raja Segaran, and Lian Pin Koh.
2019. {``Deep Learning for Environmental Conservation.''} \emph{Current
Biology} 29 (19): R977--82.
\url{https://doi.org/10.1016/j.cub.2019.08.016}.

\leavevmode\hypertarget{ref-lecun_deep_2015}{}%
LeCun, Yann, Yoshua Bengio, and Geoffrey Hinton. 2015. {``Deep
Learning.''} \emph{Nature} 521 (7553): 436--44.
\url{https://doi.org/10.1038/nature14539}.

\leavevmode\hypertarget{ref-miller2011}{}%
Miller, David A., James D. Nichols, Brett T. McClintock, Evan H.
Campbell Grant, Larissa L. Bailey, and Linda A. Weir. 2011. {``Improving
Occupancy Estimation When Two Types of Observational Error Occur:
Non-Detection and Species Misidentification.''} \emph{Ecology} 92 (7):
1422--28. https://doi.org/\url{https://doi.org/10.1890/10-1396.1}.

\leavevmode\hypertarget{ref-molinari-jobin_variation_2007}{}%
Molinari-Jobin, Anja, Fridolin Zimmermann, Andreas Ryser, Christine
Breitenmoser-Würsten, Simon Capt, Urs Breitenmoser, Paolo Molinari,
Heinrich Haller, and Roman Eyholzer. 2007. {``Variation in Diet, Prey
Selectivity and Home-Range Size of {Eurasian} Lynx {Lynx} Lynx in
{Switzerland}.''} \emph{Wildlife Biology} 13 (4): 393--405.
\url{https://doi.org/10.2981/0909-6396(2007)13\%5B393:VIDPSA\%5D2.0.CO;2}.

\leavevmode\hypertarget{ref-norouzzadeh_deep_2021}{}%
Norouzzadeh, Mohammad Sadegh, Dan Morris, Sara Beery, Neel Joshi,
Nebojsa Jojic, and Jeff Clune. 2021. {``A Deep Active Learning System
for Species Identification and Counting in Camera Trap Images.''} Edited
by Matthew Schofield. \emph{Methods in Ecology and Evolution} 12 (1):
150--61. \url{https://doi.org/10.1111/2041-210X.13504}.

\leavevmode\hypertarget{ref-NEURIPS2019_9015}{}%
Paszke, Adam, Sam Gross, Francisco Massa, Adam Lerer, James Bradbury,
Gregory Chanan, Trevor Killeen, et al. 2019. {``PyTorch: An Imperative
Style, High-Performance Deep Learning Library.''} In \emph{Advances in
Neural Information Processing Systems 32}, edited by H. Wallach, H.
Larochelle, A. Beygelzimer, F. dAlché-Buc, E. Fox, and R. Garnett,
8024--35. Curran Associates, Inc.
\url{http://papers.neurips.cc/paper/9015-pytorch-an-imperative-style-high-performance-deep-learning-library.pdf}.

\leavevmode\hypertarget{ref-Rota2016}{}%
Rota, Christopher T., Marco A. R. Ferreira, Roland W. Kays, Tavis D.
Forrester, Elizabeth L. Kalies, William J. McShea, Arielle W. Parsons,
and Joshua J. Millspaugh. 2016. {``A Multispecies Occupancy Model for
Two or More Interacting Species.''} \emph{Methods in Ecology and
Evolution} 7 (10): 1164--73.
https://doi.org/\url{https://doi.org/10.1111/2041-210X.12587}.

\leavevmode\hypertarget{ref-shao_transfer_2015}{}%
Shao, Ling, Fan Zhu, and Xuelong Li. 2015. {``Transfer {Learning} for
{Visual} {Categorization}: {A} {Survey}.''} \emph{IEEE Transactions on
Neural Networks and Learning Systems} 26 (5): 1019--34.
\url{https://doi.org/10.1109/TNNLS.2014.2330900}.

\leavevmode\hypertarget{ref-shorten_survey_2019}{}%
Shorten, Connor, and Taghi M. Khoshgoftaar. 2019. {``A Survey on {Image}
{Data} {Augmentation} for {Deep} {Learning}.''} \emph{Journal of Big
Data} 6 (1): 60. \url{https://doi.org/10.1186/s40537-019-0197-0}.

\leavevmode\hypertarget{ref-sutherland_identification_2013}{}%
Sutherland, William J., Robert P. Freckleton, H. Charles J. Godfray,
Steven R. Beissinger, Tim Benton, Duncan D. Cameron, Yohay Carmel, et
al. 2013. {``Identification of 100 Fundamental Ecological Questions.''}
Edited by David Gibson. \emph{Journal of Ecology} 101 (1): 58--67.
\url{https://doi.org/10.1111/1365-2745.12025}.

\leavevmode\hypertarget{ref-tabak_improving_2020}{}%
Tabak, Michael A., Mohammad S. Norouzzadeh, David W. Wolfson, Erica J.
Newton, Raoul K. Boughton, Jacob S. Ivan, Eric A. Odell, et al. 2020.
{``Improving the Accessibility and Transferability of Machine Learning
Algorithms for Identification of Animals in Camera Trap Images:
{Mlwic2}.''} \emph{Ecology and Evolution} 10 (19): 10374--83.
\url{https://doi.org/10.1002/ece3.6692}.

\leavevmode\hypertarget{ref-tabak_machine_2019}{}%
Tabak, Michael A., Mohammad S. Norouzzadeh, David W. Wolfson, Steven J.
Sweeney, Kurt C. Vercauteren, Nathan P. Snow, Joseph M. Halseth, et al.
2019. {``Machine Learning to Classify Animal Species in Camera Trap
Images: {Applications} in Ecology.''} Edited by Theoni Photopoulou.
\emph{Methods in Ecology and Evolution} 10 (4): 585--90.
\url{https://doi.org/10.1111/2041-210X.13120}.

\leavevmode\hypertarget{ref-vandel_distribution_2005}{}%
Vandel, Jean-Michel, and Philippe Stahl. 2005. {``Distribution Trend of
the {Eurasian} Lynx {Lynx} Lynx Populations in {France}.''}
\emph{Mammalia} 69 (2). \url{https://doi.org/10.1515/mamm.2005.013}.

\leavevmode\hypertarget{ref-andina_deep_2018}{}%
Voulodimos, Athanasios, Nikolaos Doulamis, Anastasios Doulamis, and
Eftychios Protopapadakis. 2018. {``Deep {Learning} for {Computer}
{Vision}: {A} {Brief} {Review}.''} Edited by Diego Andina.
\emph{Computational Intelligence and Neuroscience} 2018 (February):
7068349. \url{https://doi.org/10.1155/2018/7068349}.

\leavevmode\hypertarget{ref-wearn_responsible_2019}{}%
Wearn, Oliver R., Robin Freeman, and David M. P. Jacoby. 2019.
{``Responsible {AI} for Conservation.''} \emph{Nature Machine
Intelligence} 1 (2): 72--73.
\url{https://doi.org/10.1038/s42256-019-0022-7}.

\leavevmode\hypertarget{ref-weinstein_computer_2018}{}%
Weinstein, Ben G. 2018. {``A Computer Vision for Animal Ecology.''}
Edited by Laura Prugh. \emph{Journal of Animal Ecology} 87 (3): 533--45.
\url{https://doi.org/10.1111/1365-2656.12780}.

\leavevmode\hypertarget{ref-willi_identifying_2019}{}%
Willi, Marco, Ross T. Pitman, Anabelle W. Cardoso, Christina Locke,
Alexandra Swanson, Amy Boyer, Marten Veldthuis, and Lucy Fortson. 2019.
{``Identifying Animal Species in Camera Trap Images Using Deep Learning
and Citizen Science.''} Edited by Oscar Gaggiotti. \emph{Methods in
Ecology and Evolution} 10 (1): 80--91.
\url{https://doi.org/10.1111/2041-210X.13099}.

\leavevmode\hypertarget{ref-Yosinski2014}{}%
Yosinski, Jason, Jeff Clune, Yoshua Bengio, and Hod Lipson. 2014. {``How
Transferable Are Features in Deep Neural Networks?''} In
\emph{Proceedings of the 27th International Conference on Neural
Information Processing Systems - Volume 2}, 3320--28. NIPS'14.
Cambridge, MA, USA: MIT Press.

\leavevmode\hypertarget{ref-zimmermann_optimizing_2013}{}%
Zimmermann, Fridolin, Christine Breitenmoser-Würsten, Anja
Molinari-Jobin, and Urs Breitenmoser. 2013. {``Optimizing the Size of
the Area Surveyed for Monitoring a {Eurasian} Lynx ({Lynx} Lynx)
Population in the {Swiss} {Alps} by Means of Photographic
Capture-Recapture.''} \emph{Integrative Zoology} 8 (3): 232--43.
\url{https://doi.org/10.1111/1749-4877.12017}.

\end{CSLReferences}

\end{document}